\documentclass[10pt,journal,compsoc]{IEEEtran}
\ifCLASSOPTIONcompsoc
  \usepackage[nocompress]{cite}
\else
  \usepackage{cite}
\fi
\ifCLASSINFOpdf
\else
\fi
\usepackage{setspace} % This is used in the title page
\usepackage{graphicx} % This is used to load the crest in the title page
\usepackage{algorithm}
\usepackage{algorithmic}

\newcommand{\ignore}[2]{\hspace{0in}#2}

\usepackage{setspace} % This is used in the title page
\usepackage{graphicx} % This is used to load the crest in the title page
\usepackage{amsmath}
\usepackage{array}

\usepackage{url}

\usepackage{breakurl}
\usepackage[breaklinks]{hyperref}

\newcolumntype{L}[1]{>{\raggedright\let\newline\\\arraybackslash\hspace{0pt}}m{#1}}
\newcolumntype{C}[1]{>{\centering\let\newline\\\arraybackslash\hspace{0pt}}m{#1}}
\newcolumntype{R}[1]{>{\raggedleft\let\newline\\\arraybackslash\hspace{0pt}}m{#1}}

\usepackage{array}
\usepackage{graphicx}
\usepackage{booktabs}
\usepackage{mathtools, cuted}
\usepackage{multirow}
\usepackage{xcolor}
\usepackage{subcaption}
\usepackage{todonotes}
\begin{document}
\title{Multiple Linear Regression-Based Energy-Aware Resource Allocation in the Fog Computing Environment}

%==========Authors==========
\author{Ranesh~Kumar~Naha,~\IEEEmembership{Member,~IEEE,}
        Saurabh~Garg,~\IEEEmembership{Member,~IEEE,}
        Sudheer~Kumar~Battula,
        Muhammad~Bilal~Amin,~\IEEEmembership{Member,~IEEE}
        and~Dimitrios Georgakopoulos,~\IEEEmembership{Member,~IEEE}% <-this % stops a space
\IEEEcompsocitemizethanks{\IEEEcompsocthanksitem R.K. Naha, S. Garg, S.K. Battula and M.B. Amin are with the School of Technology, Environments and Design, University of Tasmania, Hobart, Tasmania,
Australia. \IEEEcompsocthanksitem D. Georgakopoulos is with the Swinburne University of Technology, Hawthorn
Victoria, Australia.
\IEEEcompsocthanksitem Corresponding Author E-mail: raneshkumar.naha@utas.edu.au, 
%\IEEEcompsocthanksitem A. Chan is with School of Engineering, University of Tasmania, Hobart, Tasmania,
}% <-this % stops an unwanted space
\thanks{Manuscript received Month XX, XXXX; revised Month XX, XXXX.}}


\markboth{Journal of \LaTeX\ Class Files,~Vol.~XX, No.~X, August~201X}%
{Naha \MakeLowercase{\textit{et al.}}: Bare Demo of IEEEtran.cls for Computer Society Journals}



\IEEEtitleabstractindextext{%
\begin{abstract}
\emph{ 
\ignore{**Introduction**} Fog computing is a promising computing paradigm for time-sensitive Internet of Things (IoT) applications. It helps to process data close to the users, in order to deliver faster processing outcomes than the Cloud; it also helps to reduce network traffic. \ignore{**Problem Statement**} The computation environment in the Fog computing is highly dynamic and most of the Fog devices are battery powered hence the chances of application failure is high which leads to delaying the application outcome. On the other hand, if we rerun the application in other devices after the failure it will not comply with time-sensitiveness. To solve this problem, we need to run applications in an energy-efficient manner which is a challenging task due to the dynamic nature of Fog computing environment. It is required to schedule application in such a way that the application should not fail due to the unavailability of energy.   
\ignore{**Objective**} In this paper, we propose a multiple linear, regression-based resource allocation mechanism to run applications in an energy-aware manner in the Fog computing environment to minimise failures due to energy constraint. Prior works lack of energy-aware application execution considering dynamism of Fog environment.
\ignore{**Methodology**} Hence, we propose A multiple linear regression-based approach which can achieve such objectives. We present a sustainable energy-aware framework and algorithm which execute applications in Fog environment in an energy-aware manner. The trade-off between energy-efficient allocation and application execution time has been investigated and shown to have a minimum negative impact on the system for energy-aware allocation.   \ignore{**evaluation/Results/Conclusion**} We compared our proposed method with existing approaches. Our proposed approach minimises the delay and processing by 20\%,  and 17\% compared with the existing one. Furthermore, SLA violation decrease by 57\% for the proposed energy-aware allocation.
}
%outperformed with the existing ones by 20\%,  and 17\% for the} in terms of delay, processing time and SLA violation.

\end{abstract}

}

\maketitle

\IEEEdisplaynontitleabstractindextext

% \fi
\IEEEpeerreviewmaketitle

\IEEEraisesectionheading{\section{Introduction}\label{sec:introduction}}

\ignore{Background}
\IEEEPARstart{F}{og} computing evolved to utilise computational resources near the users, for serving time-sensitive Internet of Things (IoT) applications  \cite{naha2020deadline}. To this end, Fog computing supports time-sensitive and real-time applications  \cite{naha2018fog} along with other applications which require edge level processing. Although Fog computing is more energy-efficient than the cloud \cite{mahmoud2018towards}, further attention is required to execute applications in the Fog environment in a time sensitive manner. Most of the Fog devices were imagined to be run by battery power and, thus, the efficient and smart use of energy will help successful application execution with fewer application failures. Energy usage for mobile devices are highly dynamic. Hence it is challenging to perform energy profiling for mobile devices. Also, energy usage is highly varied, based on the applications running on the devices, as well as the life-span of the battery and its current physical condition. Making energy profiling sustainable is another challenge since the profiling should be realistic. Additionally, energy profiling needs to be designed in a way that it could learn from the usage pattern and energy module condition. Placing applications in highly dynamic Fog devices is another competitive task, satisfying time sensitive requirements. 

\ignore{Related works}
A limited amount of research has been undertaken which specifically focused on energy-aware resource allocation in the Fog environment. In our previous works, we explored how we can place applications in the Fog environment by considering the dynamism of Fog devices and user requests \cite{naha2020deadline,naha2019multi}. Prior works have lack of addressing realistic energy-aware resource allocation in the Fog environment by considering the changing nature of resource availability in the devices. Hence, this paper aims to determine the energy-aware resource allocation goals by considering the dynamic nature of the resources in the Fog devices. The key contributions of this work are summarised as follows:
\ignore{Limitation - What needed to be solved}
\ignore{Contributions}
\begin{itemize}
%    \item Realistic energy usage profiling
    \item Energy-aware resource allocation in the highly dynamic Fog environment
    \item A sustainable solution that can find appropriate and most suitable resources based on application requirements.
\end{itemize}

\ignore{Rest of the paper-Paper Structure}
The remainder of the paper is organised as follows: Section 2 provides the background information about energy aware resource allocation research in different distributed computing paradigms. The application scenario used in our proposed solution is presented in Section 3. Detailed descriptions of the problem and the proposed solution are demonstrated in Section 4 and Section 5, respectively. In section 6, the experimental and simulation setup is presented. The experimental outcome from the simulation is analysed and discussed in section 7. Finally, Section 8 provides conclusive remarks and suggests the direction of future research.

\section{Related Work}
%novelty of the contributions
Resource allocation in an energy-aware manner is studied in various distributed computing paradigms. Some research work has been undertaken for the Fog-Cloud and Fog environments. In this section, in order to explain the novelty of our research contributions, we thoroughly analyse various work undertaken for energy-aware resource allocation.  
% Works done by others
% \subsection{Energy-aware resource allocation in Grid computing paradigm}
% \subsection{Energy-aware resource allocation in Cluster computing paradigm}
% \subsection{Energy-aware resource allocation in Cloud computing paradigm}
% \subsection{Energy-aware resource allocation in Fog-Cloud computing paradigm}
Borylo et al. \cite{borylo2016energy} proposed an energy-aware method for a wide area Software Defined Network (SDN) to provision resources dynamically in the Fog-Cloud environment. In their proposed approach, brown DCs and green DCs have been considered in which the assumption is brown DCs are contributing to $CO_2$ emission but green DCs are not contributing to $CO_2$ emission at all. Based on their assumption, requests are being handled in the Fog-Cloud environment. In their approach, energy-aware resource handling is only considered for the Cloud, while latency-aware requests are processed by the Fog. 
% \subsection{Energy-aware resource allocation in Fog computing paradigm}
An Energy-Aware Offloading Clustering Approach (EAOCA) is proposed by Bozorgchenani et al. \cite{bozorgchenani2017energy} to ensure network fairness, considering the Fog node energy level. The offloading is done in three phases. In the first phase, Fog nodes are classified by their energy availability. In the second phase, the Fog cluster has been selected by checking its capability to handle the request. Finally, the Fog cluster member is associated with the Fog cluster head in order to perform processing. Their proposed approach generated different policies by choosing the Fog nodes base power level and clustering the updating intervals during simulation. Simulation has been carried out for all the policies and the results were analysed from different generated policies. Nan et al. \cite{nan2017adaptive} proposed a computation offloading method for the Fog-Cloud environment which considered energy-awareness. The work referred to the Fog-Cloud environment as the Cloud of Things (CoT). To reduce energy emission, their framework proposed solar power as the primary power for Fog nodes. Lyapunov optimisation was employed to manage the offloading decision in order to optimise processing time and monetary costs. Their proposed system influenced the application processing in the Fog tier rather than in the Cloud tier.   

Naranjo et al. \cite{naranjo2019focan} proposed an architecture in which a device can provide services in an efficient way with low energy usage; this is known as the Fog Computing Architecture Network (FOCAN). The main goal of the FOCAN model is to demonstrate the communication between intelligent city components and services, and between Fog computing environments. In order to accommodate various accessible technologies (e.g. 3G/4 G, WiFi, ZigBees), the intelligent city is comprised of many heterogeneous elements that satisfy multiple requests from different users. FOCAN minimises the average power consumption of the Fog node and efficiently performs communication. Through two case studies, La et al. \cite{la2019enabling} suggested an approach involving device-driven and human-driven knowledge as essential to minimising energy consumption and latency in Fog computing. The first applies the learning of the computer to detect user behaviour and to adjust the Medium Access Control (MAC) layer of low latency programming for the sensor devices. They developed an algorithm for an intelligent union system in the second case study on task offloading, in the presence of multiple Fog nodes in the region, to select its task offloading decision while minimising its own energies and latency targets. The work analysed the energy and latency of end users with the number of Fog nodes. As the number of Fog nodes increased in the area, they showed substantial energy and latency reduction. Although the performance gain decreased marginally when the number of Fog nodes increased, it is obvious that the users benefit from the dense deployment of Fog nodes. 

In order to place an application module or task on Fog computing, Mahmud et al. \cite{mahmoud2018towards} proposed an energy-aware task allocation strategy. The work explored the importance of achieving reduced latency and energy consumption in the interaction between Fog computing and the two-tier Cloud of Things paradigm. In contrast to default assignments and Cloud-only policies, their simulation-based evaluation showed the efficiency of the energy-aware task allocation strategy. In contrast with the Cloud-only and Fog-default allocations, this solution was found to be more energy efficient.

The above mentioned research work deals with energy-aware application placement in the Fog environment. However, none of these focused on finding energy-aware resource allocation, considering various system characteristics, such as energy usage, CPU utilisation, power, mobility, network communication and response time. Hence, this work finds the appropriate resource by considering the energy usage of the Fog devices. This approach helps to manage the application execution in a way that minimises overheads, in terms of execution time and energy consumption.

%\section{Application Scenario}

\section{Proposed solution}
%\subsection{Problem Definition}
%How to model power profile of different Fog devices?  \\

This work is aimed to solve the following problem: How to achieve an energy-aware resource allocation objective in the Fog environment and how to make the Fog environment sustainable when satisfying time-sensitive application requirements while available resources in the devices are changing dynamically? To make the system sustainable, the system should have some intelligence to predict which resource is suitable for energy-aware resource allocation. Hence, multiple linear regression is employed to manage application execution in an energy-aware manner. Based on the data set of energy usage by the application in the Fog devices, it is possible to predict which device will best suited for energy-aware application processing. \ignore{Since the energy usage and resource availability information collected from the devices can be labeled, it is possible to employ supervised machine learning to improvise energy-awareness of the system. Unsupervised machine learning is computationally complex and Fog device has limited computational capacity, hence we choose supervised learning which computationally less complex.} An energy-aware resource allocation prediction is developed by using the linear regression-based approach. \ignore{why linear regression?} Using linear regression, we will be able to find how all the independent variables cause changes for the dependent variables. In our experimental scenario, all variables are interchangeably independent and dependent. Hence, it is necessary to quantify thoroughly the relationship between each one for energy-aware resource allocation in the Fog environment. \ignore{what are the variables?} The variables in our system are energy profiles, CPU utilisation, power, mobility, network communication and response time.

%%concepts about supervised and unsupervised learning
%%=======https://www.guru99.com/supervised-vs-unsupervised-learning.html
\subsection{Overall system design}
\begin{figure}[htbp]
	\centering
	
	\includegraphics[width=3.3in]{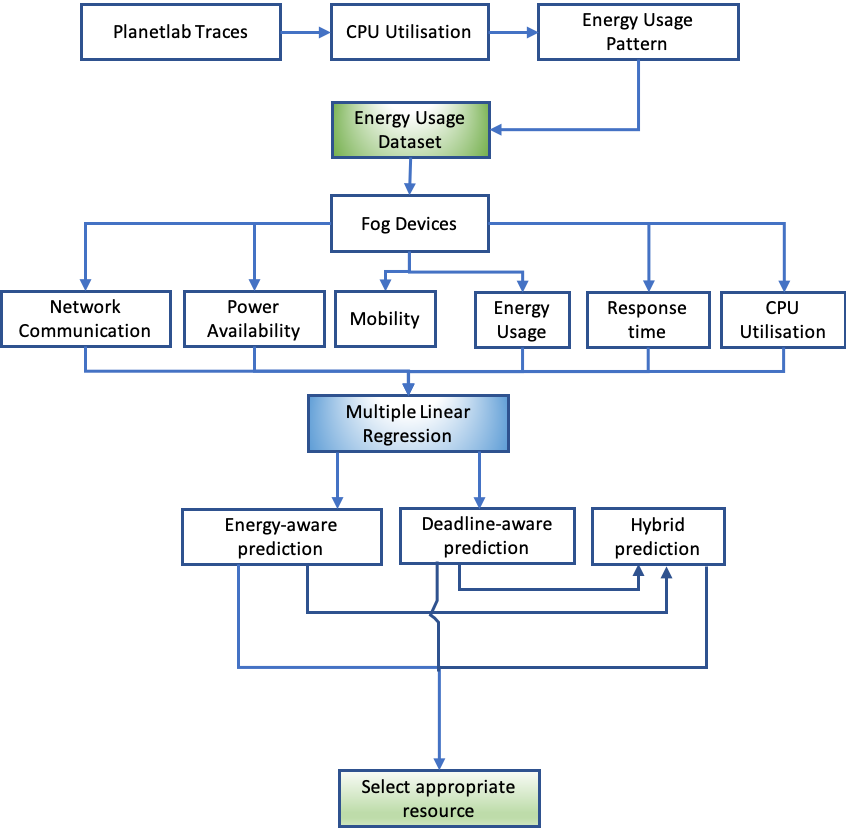}
	\caption{Overall system design.}
	\label{over}
\end{figure}

The overall goal of the proposed system is to find appropriate resources for energy-aware resource allocation. We are using planetLab workload traces to find the energy usage pattern, based on CPU utilisation in the Fog environment. \ignore{We divide the data set into training and testing dataset. Then} We applied multiple linear regression approaches to predict which resource will work better for energy-awareness, deadline-awareness and hybrid manner. \ignore{Error will be adjusted by analysing the application execution performance.}  Figure \ref{over} shows the processes in the system. \ignore{Some component are explained further in the following sections.}

Energy-aware resource allocation will only consider energy usage of Fog device while choosing best Fog device for the execution of the application. On the other hand, deadline-aware prediction basically focuses on meeting the deadline to fulfill the time-sensitiveness of the application. Then the best resource selection process was tuned up by considering how to achieve the trade-off between energy and time sensitiveness. The multiple linear regression considered Fog device characteristics, such as network communication, power availability, mobility, energy usage, response time and CPU utilisation.

\subsection{Multiple Linear Regression in Fog}
%ML techniques

%what approach
Fog application execution completion time varied on CPU utilisation, mobility, network communication and response time. Here the application execution time is the dependent variable which depends on four other dependent variables such as: CPU utilisation, mobility, network communication and response time. Hence, we can predict application execution time with the following multiple linear regression equation.

\begin{equation}
    \begin{split}
    \label{eq1}
        ETP_{a_x} = \beta_0 + \beta_1CPU_{u_{d_r}} +  \beta_2Dev_{m_{d_r}} +  \beta_3Net_{comm_{d_r}} \\
        +  \beta_4Res_{t_{d_r}} + \epsilon
    \end{split}
\end{equation}

%why
\textcolor{black}{We have chosen multiple linear regression because prediction of application completion time from multiple quantitative independent variable is possible with this regression method. Simple regression is not suitable because we have multiple independent variables.}   

In the above equation, $ETP_{a_x}$ is the predicted execution time for the application $x$. $\beta_0$ is the intercept of the model which represents its value, while all dependent variable are equal to zero. \ignore{Throughout the experiments we have to find which value is most suitable for our scenario.} $\beta_1CPU_{u_{d_r}}$ is the CPU utilisation for device $r$ for application $x$. $\beta_2Dev_{m_{d_r}}$ is the device mobility for device $r$ for application $x$. Similarly, $\beta_3Net_{comm_{d_r}}$ and $\beta_4Res_{t_{d_r}}$ are the network communication and response times for device $r$ and for application $x$, respectively. $\beta_1$, $\beta_2$, $\beta_3$ and $\beta_4$ are the coefficients of the model which determine the slope of the regression line and which describe the proposed model. $\epsilon$ is the error rate which defines the difference between the proposed regression model and actual observation results.

On the other hand, power availability influences whether it is possible to complete an intended task with the particular Fog resources or not. Hence, power availability will be considered as a dichotomous predictor variable, defining whether it is positive or negative. Furthermore, the energy usage patterns of the application and the devices (also known as power profiling) have direct impact on energy-aware resource allocation. After considering power availability and energy usage patterns, the regression model will be as follows:

\begin{equation}
    \begin{split}
    \label{eq2}
        ETP_{a_x} = \beta_0 + \beta_1CPU_{u_{d_r}} +  \beta_2Dev_{m_{d_r}} +  \beta_3Net_{comm_{d_r}} + \\ \beta_4Res_{t_{d_r}} + 
        \beta_5Pow_{avail_{d_r}} +
        \beta_6Ene_{usage_{d_r}} +
        \epsilon
    \end{split}
\end{equation}

Where, $\beta_5Pow_{avail_{d_r}}$ and $\beta_6Ene_{usage_{d_r}}$ are the power availability and energy usage pattern in the device $r$ for the application $x$ respectively. Equation  \ref{eq2} is most suitable for a deadline-aware application. However, the following regression model will be used for energy-aware allocation.  

\begin{equation}
    \begin{split}
    \label{eq3}
        EEC_{a_x} = \beta_0 + \beta_1CPU_{u_{d_r}} +  \beta_2Dev_{m_{d_r}} +  \beta_3Net_{comm_{d_r}} +  \\
        \beta_4Res_{t_{d_r}} + 
        \beta_5Pow_{avail_{d_r}} +
        \beta_6AE_{time{d_r}} +
        \epsilon
    \end{split}
\end{equation}
In the above equation, $EEC_{a_x}$ is the predicted energy consumption for the application $x$. $AE_{t_{d_r}}$ is the application execution time for the $x$ application in the device $d_r$. For both $ETP_{a_x}$ and $EEC_{a_x}$ the best device will be selected using the following equations (Equation \ref{eq4} and Equation \ref{eq5}).

\begin{equation}
    \begin{split}
    \label{eq4}
        {a_x}_{srd} = Min\Big({ETP_{a_x}}_1, {ETP_{a_x}}_2,  {ETP_{a_x}}_3  \ldots  {ETP_{a_x}}_n\Big) 
    \end{split}
\end{equation}

\begin{equation}
    \begin{split}
    \label{eq5}
        {a_x}_{sre} = Min \Big( {EEC_{a_x}}_1, {EEC_{a_x}}_2,  {EEC_{a_x}}_3  \ldots  {EEC_{a_x}}_n \Big)
    \end{split}
\end{equation}

${a_x}_{srd}$ is the selected resource which best meets the deadline and ${a_x}_{sre}$ is the
selected resource which best meets energy awareness.

% \begin{equation}
% \label{eqnor}
%     \mu_{mN}(x)= \begin{cases} 
%   0, &  (x < a) \ or \ (x > d) \\
%   \frac{x-a}{b-a},   & a \leq x \leq b \\
%   1,       & b \leq x \leq c \\
%   \frac{d-x}{d-c},   & c \leq x \leq d
%   \end{cases}
% \end{equation}

% \begin{equation}
% \label{eqhigh}
%     \mu_{mH}(x)= \begin{cases} 
%   0, &  x < a \\
%   \frac{x-a}{b-a},   & a \leq x \leq b \\
%   1,       & x > b
%   \end{cases}
% \end{equation}

% \begin{figure}[htbp]
% 	\centering
% 	\label{fig_ufmrp}
% 	\includegraphics[width=3in]{UFmrp.png}
% 	\caption{Membership for mrp score.}
% \end{figure} 

\subsection{Modelling Energy Usage of Fog Device}
%\subsection{Dataset}
We used the Planetlab workload dataset \cite{beloglazov2012energy} for modelling energy usage patterns for Fog devices. 
Planetlab has a CPU utilisation dataset; based on that we can calculate the energy consumption pattern of the node. We used a similar approach to calculate the energy consumption of the Fog node since our goal is energy-aware resource allocation. The Planetlab dataset contains CPU utilisation data for every five minutes from thousands of virtual machines. We choose CPU utilisation data randomly from the dataset for each Fog device in order to find the energy usage pattern.   

\subsection{Proposed algorithm}
Energy and deadline-aware resource allocation is presented in Algorithm \ref{alg1}. The proposed algorithm find the best Fog resource for application execution, using multiple linear regression. Based on the application requirements, the proposed algorithm finds a resource which can execute the application in a deadline-aware, energy-aware or hybrid manner. 

\begin{algorithm}[htbp]
	\caption{Energy and deadline-aware resource allocation.}
	\label{alg1}
	\footnotesize
%	\begin{singlespace}
	\hspace*{\algorithmicindent} \textbf{Input: $ FD[id, CPU_{util}, Dev_{mob}, Net_{comm}, Res_{time}, Pow_{avail}, \\ Ene_{usage}, AE_{time}]$, $a_x$}  \\
	\hspace*{\algorithmicindent} \textbf{Output:$FD_{id}$} 
	\begin{algorithmic} 
		%\STATE $P_{min} \leftarrow c$
		%\STATE $B_{min}\leftarrow c$
		%\STATE $R \leftarrow ResourceList<instances>$
		\STATE $\epsilon \leftarrow 0$;
		\FORALL{$FD[id]$}
		\STATE Calculate ${ETP_{app}}_x$ 
		\STATE Calculate ${EEC_{app}}_x$
		\ENDFOR
		\IF {$Req = D_{aw}$} %deadline aware
		\STATE Calculate ${a_x}_{srd}$
		\RETURN $FD_{id}$
        \ELSIF {$Req = E_{aw}$} %energy aware
        \STATE Calculate ${a_x}_{sre}$
        \RETURN $FD_{id}$
        \ELSIF {$Req = DE_{aw}$} %deadline and energy aware
        \STATE $B_r \leftarrow FD[]$ %select best resource considering both energy and deadline
	    \RETURN $FD_{id}$
	    \ELSE
	    \RETURN $NULL$
	   \ENDIF
	\end{algorithmic}
%	\end{singlespace}
\end{algorithm}

The input of the algorithm is the list of Fog devices and the application requests. Each Fog device has its id along with its resource characteristics, such as CPU utilisation, mobility, network communication, response time, power availability, energy usage and application execution performance. The output of the algorithm is the best suited Fog device for the submitted application. The error for the regression model will be set to 0 initially. Then, execution time and energy consumption will be estimated for the requested application for each Fog device. After that, based on user requirement, the algorithm will select the best resource for the requested application. While selecting the best resource, the algorithm will consider either time-sensitiveness or energy awareness. It also considers the trade-off between both energy and deadline.     

% \section{AI and Machine Learning}

% Regression: Predicting best suitable resource based on device mobility, native resource usage and configuration, Algorithm name- Linear regression

% Clustering: Fog cluster segmentation - Automatic cluster forming using Fog devices and select master Fog node, Algorithm name- K-means, LDA

% Collaborative Filtering: Recommending Fog device which will perform better for Fog processing (Recommender system), Algorithm name- Alternating Least Squares (ALS)

% Dimensionality reduction: Reducing the number of redundant features/variables. 1) selecting only important features during selecting Fog devices for processing (For example for a specific scenario device configurations are similar but there is difference in utilization so we do not need to analyze Fog device configuration feature). 2) Removing redundant features e.g. Mbps and Gbps are linearly dependent, Algorithm name- Principal component Analysis (PCA)

%reference: https://www.youtube.com/watch?v=ggIk08PNcBo
%see following resources
%https://www.youtube.com/watch?v=vof2vhfqoBo
%https://towardsdatascience.com/linear-regression-in-real-life-4a78d7159f16

% \subsection{Energy Model}
% \subsection{Real traffic model}

\subsection{Performance Metrics}
% Trade-off between energy and cost
% Network usage (Mahmoud)
% Energy Consumption
% End to End Latency
% Deadline-aware allocation
% Energy-aware allocation

\textcolor{black}{All the performance metrics adopted from our previous works \cite{naha2019deadline,naha2019multi}.
\\ \\
\textbf{Delay:} We considered delay between the user and the Fog resources. Delay is the time between task submission and starting task execution. It can be calculated as follows:}

%\vspace{-2ex}
\begin{equation}
%\begin{aligned}
\begin{split}
\label{eq1}
\color{black}
d_{t}^x=E_{st}^x-U_S 
\end{split}
%\end{aligned}
\end{equation}

\textcolor{black}{In Equation \ref{eq1}, $d_t^x$ denotes the delay for the $x$ Fog device which is involved in task execution. $E_{st}$ is the task start time and $x-U_S$ is the time when the user requested for the task execution.}
\\
\\
\textcolor{black}{\textbf{Processing time:} Processing time is the required time to process a task. It is the time between task processing start time $p_{st}$ and task processing end time $p_{en}$ which can be calculated by using the Equation \ref{eq2}.}

%\vspace{-2ex}
\begin{equation}
%\begin{aligned}
\begin{split}
\label{eq2}
\color{black}
Pt_t^x=p_{en}^x-p_{st}^x
\end{split}
%\end{aligned}
\end{equation}

\textcolor{black}{In the above equation $x$ is the Fog device which is involved in task execution and $Pt_t$ is the processing time for task $t$. }
\\ \\
\textcolor{black}{\textbf{Processing Cost:} We considered connectivity and messaging costs for processing costs. These costs are based on the AWS IoT pricing model. Cost is from \$1 to \$1.65 per million messages for messaging and from \$0.08 to \$0.132 for connectivity cost for per million minutes for various regions. We considered the price that has been allocated to the Sydney region. Processing cost can be calculated as follows:}

%\vspace{-2ex}
\begin{equation}
%\begin{aligned}
\begin{split}
\label{eq2}
\color{black}
Pc_t=\sum_{k=a}^{n}(M_c + C_c)
\end{split}
%\end{aligned}
\end{equation}

\textcolor{black}{In the above equation, $M_c$ is the messaging cost, $C_c$ is the connectivity cost and $Pc_t$ is the total processing cost. We calculated the cost for Fog device $a$ to Fog device $n$.}

\textbf{Service Level Agreement (SLA): }Service quality is generally guaranteed by the SLA. The provider is responsible for the maintenance of an adequate response time to avoid the violation of SLA. We will measure the response time and cost as agreed by the SLA. In the Fog, users' dynamic requirements will be response time or cost. If the provider is unable to serve according to the agreed requests of the users, then the provider will have to pay for the violation. An SLA violation penalty will follow a linear function which is similar to other related works of \cite{yeo2005service,rana2008managing,irwin2004balancing,wu2011sla,wu2014sla}. The function is as follows:

 %in which the user is deducted from the total usage bill of the users
 
\begin{equation}
Penalty = \alpha+\beta \times DT 
\end{equation}
where, $\alpha$ is a constant value for the penalty, $\beta$ is the penalty rate and $DT$ is the delay time. Delay time is the extra time that users waited as stated in the SLA for obtaining a response. The percentage of SLA violations is also calculated.
\section{Experimental Setup and Simulation Parameters}

% Modeling power? Battery drainage

% Modeling CPU utilization? Native utilization modeling\\
% Autoregressive Integrated Moving Average (ARIMA)\\
% Sample utilization model for: Mobile phone, laptop, tablet, PC (Do we need to collect?) %
% https://github.com/IrinaMax/CPU-usage-and-anomaly-detection
% \\
% https://people.duke.edu/~rnau/411arim2.htm \\
% \\ \\
% Modeling application response time? Based on network connectivity, location and processing\\
% Modeling network communication? Based on network connectivity \\

\subsection{Experimental Setup}
%F2
To control over the experimental environment, we chose simulation for the evaluation of the proposed method. We adopted a simulation environment and performance parameters from our previous works \cite{naha2019deadline} \cite{naha2019multi}. In addition, we modelled a realistic Fog environment using the CloudSim \cite{calheiros2011cloudsim} toolkit, similar to our previous work \cite{naha2019deadline} \cite{naha2019multi}. All submitted tasks followed deadlines which varied dynamically from 10\% to 80\%. Successful execution of the application by maintaining deadlines indicated successful processing. \ignore{from F1Jour} Based on the previous literature \cite{mahmud2018quality,skarlat2017towards}, we tested the proposed method by increasing the number of application submissions. Hence, 70 to 560 applications have been submitted to the Fog environment, increasing by 70 applications each time. 

\subsection{Simulation Parameters}
Table \ref{tsFogSimPara} illustrates the parameters used for the simulation. Table \ref{DybFogSimPara2} represents the other parameters that are used to model dynamic user behaviour, distance, battery life and CPU availability fluctuations.

\begin{table}[htbp]
	\centering
	\footnotesize
	\caption{Simulation Parameters}
	\label{tsFogSimPara}
	%\begin{tabular}{|c|l|l|l|} \hline
	%\begin{tabular}{|p{1cm}|p{2.2cm}|p{2cm}|p{1.7cm}|} \hline
	\begin{tabular}{L{5cm}|L{2.5cm}} 
		\toprule[1pt]
		\textbf{Parameter} & \textbf{Value} \\ \hline
		
        \textbf{Fog Server Configuration} & \\ \hline
		MIPS (Millions Instruction Per Second) & 10000  \\ \hline
		No of Pes & 1 \\ \hline
		No of Host & 1 \\ \hline
		Bandwidth (bps) & 1000000 \\ \hline
		RAM & 302768 \\ \hline
		
		\textbf{Fog Device Configuration} & \\ \hline
		MIPS (Millions Instruction Per Second) & 2000 to 6000  \\ \hline
		No of Pes & 1 \\ \hline
		No of Host & 1 \\ \hline
		Bandwidth (bps) & 100000 \\ \hline
		RAM & 2048 \\ \hline
		
		\textbf{Task Configuration} & \\ \hline
		Task Length (MI) & 3000  \\ \hline
		Data Size & 5120 and above \\ \hline
		
		\textbf{Sub Task Configuration} & \\ \hline
		Task Length (MI) & 500  \\ \hline
		Data Size & 5120 and above \\ \hline
		\bottomrule[1pt]
		
	\end{tabular}
\end{table}

\begin{table}[htbp]
	\centering
	\footnotesize
	\caption{Other Parameters}
	\label{DybFogSimPara2}
	%\begin{tabular}{|c|l|l|l|} \hline
	%\begin{tabular}{|p{1cm}|p{2.2cm}|p{2cm}|p{1.7cm}|} \hline
	\begin{tabular}{L{5cm}|L{2.5cm}} 
		\toprule[1pt]
		\textbf{Parameter} & \textbf{Value} \\ \hline
		\textbf{No of Task per App} & 10\\ \hline
		\textbf{Minimum deadline for tasks} & 4\\ \hline
		\textbf{CPU availability fluctuation} & 50\% - 130\%\\ \hline
		\textbf{Distance} & 5 to 40 Meter\\ \hline
		\textbf{Battery power} & 20\% to 90\% \\\hline
        \textbf{CPU Utilisation variation during task execution} & 10\% to 40\%         \\
        \bottomrule[1pt]
		
	\end{tabular}
\end{table}

%70 to 560 applications have been submitted to the Fog environment by increasing 70 applications each time to test the performance of the proposed MC-Based policy with QoE-aware policy with the increasing number of application requests.

The main goal of this work is to allocate the application tasks to the Fog infrastructure\ignore{, and not to the Cloud}. During the simulation, different evaluation scenarios were followed. In the first evaluation scenario, 70 applications were submitted to the Fog environment in the initial stage, then the number of application submissions was increased gradually, up to 560 applications. We measured delay, processing time and cost for this evaluation scenario. In the second evaluation scenario, with the increased number of application submissions, SLA violation was measured with and without reservation. In the last evaluation scenario, each dynamic parameter for users and devices was varied. These dynamic parameters were variations of user deadlines, free resources, battery power and CPU utilisation fluctuation. 

%\subsection{Simulation Scenarios}

\section{Results and Discussion}
\begin{figure}[htbp]
        \includegraphics[width=0.48\textwidth, height=5.3cm]{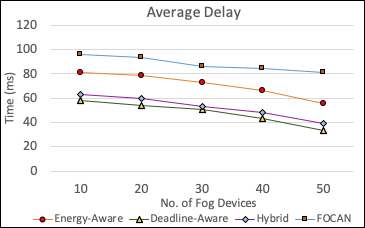}
        \caption{Average delay when the number of Fog devices is changing.}
        \label{FDC1}
\end{figure}

\begin{figure}[htbp]
        \includegraphics[width=0.48\textwidth, height=5.3cm]{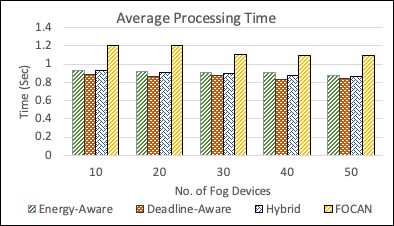}
        \caption{Average processing time when the the number of Fog devices is changing.}
        \label{FDC2}
\end{figure}

\begin{figure}[htbp]
        \includegraphics[width=0.48\textwidth, height=5.3cm]{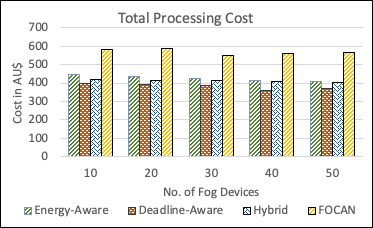}
        \caption{Total processing cost when the number of Fog devices is changing.}
        \label{FDC3}
\end{figure}

\begin{figure}[htbp]  
        \includegraphics[width=0.48\textwidth, height=5.3cm]{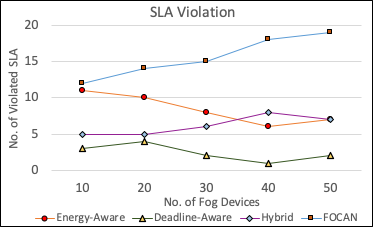}
        \caption{SLA violation when the number of Fog devices is changing.}
        \label{FDC4}
\end{figure}

Energy-aware, deadline-aware and hybrid resource allocation methods were compared with FOCAN \cite{naranjo2019focan}. We measured average delay, average processing time, total processing time and SLA violation for all methods. Figures \ref{FDC1}, \ref{FDC2}, \ref{FDC3} and \ref{FDC4} show average delay, average processing time, total processing time and SLA violation respectively, while a number of Fog devices are changing in the Fog environment. According to the Figure \ref{FDC1}, deadline-aware, energy-aware and hybrid resource allocation methods performed better, compared with FOCAN. On average, the improvements were 46\% for deadline-aware, 20\% for energy-aware and 41\% for the hybrid method, compared with FOCAN. However, the highest improvement was 59\% when 50 Fog devices are active in the application environment. On the other hand, the improvement was  39\% when the minimum number of Fog devices was in operation in the Fog environment. From the experiments it was found that average delay is decreasing with the increasing number of Fog resources. 

%found for deadline-aware resource allocation compared with FOCAN.

Average processing time improved by 20\% to 24\% for energy-aware, deadline-aware and hybrid resource allocation methods, compared with FOCAN as shown in Figure \ref{FDC2}. It was observed that the average processing time is lower when a higher number of Fog devices is in operation in the Fog environment.

Compared with FOCAN, the total processing cost was lower in the proposed methods. Total costs were 33\% lower for deadline-aware, 25\% lower for energy-aware and 28\% lower for the hybrid method compared with FOCAN, as shown in Figure \ref{FDC3}. It was observed that the processing cost is less when the higher number of Fog devices is in operation in the Fog environment.

SLA violation decreased by 83\% for deadline-aware, 60\% lower for energy-aware and 43\% lower for hybrid method on average compared with FOCAN as shown in Figure \ref{FDC4}. Proposed methods were outperformed in terms of SLA violation compared with FOCAN. However, it was observed that SLA violation is lower when the number of available Fog devices is higher in the Fog environment.

%  \caption{Impact on delay, processing time, cost and SLA violation while number of Fog devices are changing.}\label{IRP}

%Four figures adjustment
% \begin{figure*}[t]
%     \centering
%     \begin{subfigure}[b]{0.48\textwidth}
%         \includegraphics[width=\textwidth, height=5.3cm]{FDC1_avgdelay.png}
%         \caption{Average Delay}
%         \label{IRP_TT}
%     \end{subfigure}
%     \begin{subfigure}[b]{0.48\textwidth}
%         \includegraphics[width=\textwidth, height=5.3cm]{FDC2_avgprotime.png}
%         \caption{Average Processing Time}
%         \label{IRP_NC}
%     \end{subfigure}
%     \begin{subfigure}[b]{0.48\textwidth}
%         \includegraphics[width=\textwidth, height=5.3cm]{FDC3_tpprocost.png}
%         \caption{Total Processing Cost}
%         \label{IRP_NConv}
%     \end{subfigure}
%     \begin{subfigure}[b]{0.48\textwidth}
%         \includegraphics[width=\textwidth, height=5.3cm]{FDC4_SLAV.png}
%         \caption{SLA Violation}
%         \label{IRP_BF}
%     \end{subfigure}
%   \caption{Impact on delay, processing time, cost and SLA violation while number of Fog devices are changing.}\label{IRP}
% \end{figure*}

\begin{figure}[htbp]
        \includegraphics[width=0.48\textwidth, height=5.3cm]{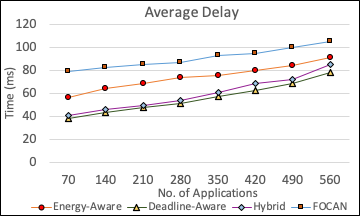}
        \caption{Average delay when the number of application submissions is changing.}
        \label{APP1}
\end{figure}

\begin{figure}[htbp]
        \includegraphics[width=0.48\textwidth, height=5.3cm]{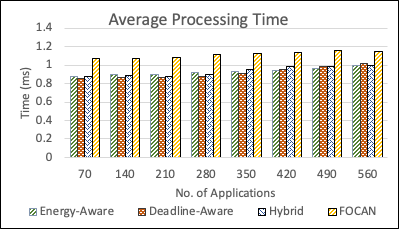}
        \caption{Average processing time when the number of application submissions is changing.}
        \label{APP2}
\end{figure}

\begin{figure}[htbp]    
        \includegraphics[width=0.48\textwidth, height=5.3cm]{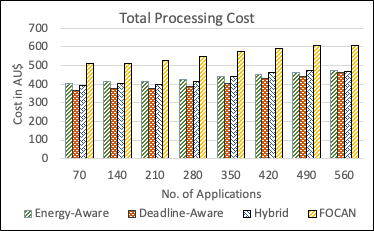}
        \caption{Total processing cost when the number of application submissions is changing.}
        \label{APP3}
\end{figure}

\begin{figure}[htbp]    
        \includegraphics[width=0.48\textwidth, height=5.3cm]{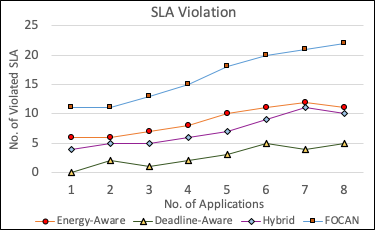}
        \caption{SLA violation when the number of application submissions is changing.}
        \label{APP4}
\end{figure}

Figure \ref{APP1}, \ref{APP2}, \ref{APP3} and \ref{APP4} show average delay, average processing time, total processing time and SLA violation respectively when the number of application submission is changing in the Fog environment. According to the Figure \ref{APP1}, deadline-aware, energy-aware and hybrid resource allocation methods performed better, compared with FOCAN. On average the improvement were 39\% for deadline-aware, 35\% for energy-aware and 18\% for the hybrid method, compared with FOCAN. The best improvement were found in deadline-aware but if we consider energy-awareness, it is necessary to utilise the energy-aware method, although the improvement was less compared with deadline-aware and hybrid methods. From the experiments it was observed that the average delay is increasing with the increasing number of application submissions in the Fog environment.

Average processing time was improved by around 17\% for the proposed method, compared with FOCAN as shown in Figure \ref{APP2}. For the energy-aware method, the average processing time improvement was 14\% to 18\%. For the deadline-aware method the average processing time improvement was 11\% to 20\%. A similar processing time improvement pattern observed for the proposed hybrid method. The processing time increased with the increasing number of application submission. 

In the proposed method, the total processing cost decreased by 22\% to 27\%, as shown in Figure \ref{APP3}. However, the processing cost was increased with the increasing number of application submissions. We found similar costs for both energy-aware and hybrid methods.

SLA violation decreased by 85\% for deadline-aware, 57\% lower for energy-aware and 46\% lower for hybrid methods on average, compared with FOCAN as shown in Figure \ref{APP4}. The proposed methods outperformed in terms of SLA violation, compared with FOCAN, similar to the change in the number of Fog devices scenario. However, it was observed that SLA violation is lower when a lower number of applications was submitted in the Fog environment.

From our experimental evaluation we found that a higher number of available Fog devices minimises the risk of SLA violation. Since the Fog computing environment is best for time sensitive applications, it is required to have a lower number of SLA violations for successful application execution. We found that the hybrid method is better for an energy-efficient application environment, although delay and processing time are slightly higher.
    
%\caption{Impact on delay, processing time, cost and SLA violation while number of application submission is changing.}\label{IRP}

% %Four figures adjustment
% \begin{figure*}[t]
%     \centering
%     \begin{subfigure}[b]{0.48\textwidth}
%         \includegraphics[width=\textwidth, height=5.3cm]{App1_avgdelay.png}
%         \caption{Average Delay}
%         \label{IRP_TT}
%     \end{subfigure}
%     \begin{subfigure}[b]{0.48\textwidth}
%         \includegraphics[width=\textwidth, height=5.3cm]{App2_avgprotime.png}
%         \caption{Average Processing Time}
%         \label{IRP_NC}
%     \end{subfigure}
%     \begin{subfigure}[b]{0.48\textwidth}
%         \includegraphics[width=\textwidth, height=5.3cm]{App3_tpprocost.png}
%         \caption{Total Processing Cost}
%         \label{IRP_NConv}
%     \end{subfigure}
%     \begin{subfigure}[b]{0.48\textwidth}
%         \includegraphics[width=\textwidth, height=5.3cm]{App4_SLAV.png}
%         \caption{SLA Violation}
%         \label{IRP_BF}
%     \end{subfigure}
%   \caption{Impact on delay, processing time, cost and SLA violation while number of application submission is changing.}\label{IRP}
% \end{figure*}

\section{Conclusion}
The devices in the Fog environment are mostly run on battery power. Hence, energy-aware resource allocation is required for successful application execution. In this work, we proposed a resource allocation method which considers energy and deadline while selecting resources for application execution. We employed multiple linear regression for selecting more appropriate resources, based on user requests. Since the energy-aware method did not perform well for some cases, we further developed a hybrid method which considers both deadline and energy-awareness. The proposed method performed better than the existing one. In the future, we will implement the proposed resource allocation mechanism in an operational Fog environment. 

\ignore{problem summary}

\ignore{Significance of the problem} 

\ignore{proposed approach} 

\ignore{Key findings}

\ignore{Future works}

\bibliographystyle{IEEEtran}
\bibliography{paper}
\end{document}